\documentclass{aa}  
\usepackage{rotating}
\usepackage{graphicx}
\usepackage{txfonts}
\usepackage{natbib}
\bibpunct{(}{)}{;}{a}{}{,} 

\begin{document}

\title{3D Spectroscopy with VLT/GIRAFFE - IV: Angular Momentum \&
Dynamical Support of Intermediate Redshift Galaxies\thanks{Based on
FLAMES/GIRAFFE Paris Observatory Guaranteed Time Observations
collected at the European Southern Observatory, Paranal, Chile, ESO
No. 71.A-0322(A) and 72.A-0169(A)}}

\author{M. Puech\inst{1}, F. Hammer\inst{1}, M. D. Lehnert\inst{2},
\and H. Flores\inst{1}}

\authorrunning{Puech et al.}

\titlerunning{Angular Momentum of Intermediate Redshift Galaxies}

\offprints{mathieu.puech@obspm.fr}

\institute{Laboratoire Galaxies Etoiles Physique et
        Instrumentation, Observatoire de Paris, 5 place Jules Janssen,
        92195 Meudon France\\
      \and
        Max-Planck-Institut f\"ur extraterrestrische Physik,
        Giessenbachstra\ss e, D-85748 Garching, Germany}

\date{Received ............... ; accepted ............... }

\abstract{One of the most outstanding problems related to numerical
models of galaxy formation is the so-called ``angular momentum
catastrophe'', i.e., the inability to explain theoretically the high
angular momentum observed in local disk galaxies.} {We study the
evolution of the angular momentum from z$\sim$0.6 to z=0 to further
our understanding of the mechanisms responsible for the large angular
momenta of disk galaxies observed today. This study is based on a
complete sample of 32, 0.4$\leq$z$\leq$0.75 galaxies observed with
FLAMES/GIRAFFE at the VLT. Their kinematics had been classified as
rotating disks (11 galaxies), perturbed rotators (7 galaxies), or
complex kinematics (14 galaxies).} {We have computed the specific
angular momentum of disks ($j_{disk}$) and the dynamical support of
rotating disks through the $V/\sigma$ ratio. To study how angular
momentum can be acquired dynamically, we have compared the properties
of distant and local galaxies, as a function of their kinematical
class.} {We find that distant rotating disks have essentially the same
properties ($j_{disk}$ and $R_d$) as local disks, while distant
galaxies with more complex kinematics have a significantly higher
scatter in the $j_{disk}$--$V_{max}$ and $R_{d}$--$V_{max}$ planes. On
average, distant galaxies show lower values of $V/\sigma$ than local
galaxies, the lowest $V/\sigma$ values being reached by distant
galaxies showing perturbed rotation. This can probably be attributed
to heating mechanisms at work in distant disks.} {We found
observational evidence for a non-linear random walk evolution of the
angular momentum in galaxies during the last 8 Gyr. The evolution
related to galaxies with complex kinematics can be attributed to
mergers, but not to (smooth) gas accretion alone. If galaxies observed
at intermediate redshift are related to present-day spirals, then our
results fit quite well with the ``spiral rebuilding'' scenario
proposed by Hammer et al. (2005).}

\keywords{Galaxies: evolution; Galaxies: formation; Galaxies: kinematics
and dynamics; Angular momentum; 3D spectroscopy.}

\maketitle

\section{Introduction}

The evolution and the origin of the Tully-Fisher (T-F) relation is
still a matter of intense debate. A strong evolution both in slope and
dispersion has been found in B band \citep[e.g.,][]{ziegler02,
boehm04}, from z$\sim$1 to z=0. More recently, \citet{conselice05}
derived the T-F relation in K band, K-band absolute magnitude being a
better tracer of stellar mass, and found such a large dispersion that
one could wonder whether or not the T-F relation even exists at
z$\sim$1. \citet[][hereafter Paper I]{flores06} derived the first T-F
relation for distant galaxies using integral field spectroscopy. The
two dimensional spatial coverage allowed them to properly identify the
dynamical nature of distant galaxies, and establish that as much as
$\sim$40\% of field galaxies are not in equilibrium, i.e., galaxies
that are not suitable to establish a proper T-F relation. They then
derived a T-F relation that does not appear to have evolved since
z$\sim$0.6, in the M$_K$ (or stellar mass) versus $V_{max}$ plane. On
the other hand, the stellar mass density increases from z=1 to z=0
\citep[e.g.,][]{drory05}: assuming a rough 30\% increase in stellar
mass from z=0.6 to z=0, we should then see a $\sim$0.1 dex shift along
the $M_*$ axis between the local and distant T-F relations. However,
one should keep in mind that typical uncertainties usually associated
with $M_*$ are of the same order, i.e., 0.1-0.2 dex (see Paper I).
Another possibility is to assume some gas accretion from the
intergalactic medium which would be directly converted into stellar
mass: while $M_*$ increases, $V_{max}$ could then increase in the same
time, because the total mass of the system mainly depends on the
rotational velocity. Then, both quantities could evolve such as the
resulting evolution in the T-F plane operates \emph{along} the
relation \citep[e.g.,][]{portinari06}. More statistics and/or future
studies of the T-F relation at higher redshifts (where the stellar
mass density was much smaller than at z=0.6) should provide a decisive
test of this point. Whether the T-F relation evolves with time will no
doubt provide important clues about relationship between the growth of
mass and the characteristics of the stellar populations, and,
ultimately, about galaxy formation and evolution in general.

The ideas concerning the origin of the T-F relation can be divided into
two broad categories. In the first one, the T-F relation originates
from the cosmological equivalence between the halo mass and the circular
velocity \citep[e.g.,][]{mo98}. The relation then comes from the fact that
the finite age of the Universe imposes a maximal radius from where matter
can be accreted to form a galaxy. The second broad categories of models
invoked to explain the T-F relation is self-regulated star formation
in disks of different mass \citep[e.g.,][]{silk97}. However, numerical
simulations taking into account both ingredients of gas accretion and
self-regulation have not been able to reproduce all aspects of the T-F
relation, such as the zero point \citep[e.g.,][]{steinmetz99}. Many
authors suggested that feedback from massive star formation or active
galactic nuclei could help to solve these discrepancies \citep[see,
e.g.,][]{eisenstein96, heavens99, steinmetz99}.

Feedback has also been suggested to solve the so-called ``angular
momentum catastrophe'' of the $\Lambda$-CDM model, i.e., the inability of
simulations to reproduce disk galaxies with sufficient angular momentum in
comparison with what is observed \citep[e.g.,][]{steinmetz99}. Feedback
has been proposed as a potential solution to this problem, which has
been investigated in detail \citep[e.g.,][]{maller02a, d'Onghia06,
governato06}.  State-of-the-art numerical simulations, including
the effects of AGN feedback driven through accretion of gas onto a
super-massive central black hole, show how a disk can re-form after
the merging of two rotating gas-rich disks, with a sufficient and
consistent amount of angular momentum \citep{robertson05}.  

The mechanisms through which galaxies may have acquired their angular
momentum has been discussed for many decades
\citep[e.g.,][]{stromberg34, hoyle49, mestel63}. The so-called
``gravitational instability paradigm'', independent of the details of
the cosmology, predicts that the angular momentum of a protogalaxy
should grow linearly with time due to tidal torques from interactions
with neighboring structures, until it decouples from the Hubble flow
\citep{peebles69, white84}. In a more modern picture, galaxies form
from infalling baryonic gas embedded into dark matter haloes, and
their angular momentum is then inherited from the halo \citep{white78,
fall80, barnes87}. After the end of the epoch during which tidal
torquing is effective, subsequent evolution of the angular momentum
takes place non-linearly, through a random walk process associated
with mergers events and/or mass accretion \citep{vitvitska02,
maller02a, peirani04}. This random walk leads to a change in the
angular momentum of the haloes, with a more significant change
(increase or decrease, depending on the geometry of the merger) during
major mergers \citep{vitvitska02, peirani04}. For a given galaxy, it
is in this way that major mergers are the main source of either
positive or negative change in its angular momentum.

Relating the angular momenta of the halo and the disk is not at
all straightforward. A reasonable assumption often made is that the
specific angular momentum $j$ (i.e., the angular momentum per unit
mass), is conserved during the collapse of the gas \citep{mestel63,
fall80}. If we also assume that both gas and dark matter are well mixed
in the proto-galaxy \citep{fall80}, then this leads to $j_{disk}\sim
j_{halo}$. Making this assumption allows models and simulations to
reproduce several properties of local disk galaxies \citep[e.g.,][]{mo98,
syer99, vandenBosch01, vandenBosch02a, tonini06}. However, some problems
remain, such as the ``mismatch of angular momentum profiles'' between the
dark matter and the disk \citep{bullock01, vandenBosch02a, maller02a},
and the inability of models to simultaneously match characteristics
such as the slope and zero-point of the T-F relation, slope and
zero-point of the radius-luminosity relation, the luminosity function
of spirals with reasonable values for the disk masses, halo structural
parameters, and circular velocity relative to virial velocity \citep[see,
e.g.,][]{dutton06}.

The build-up of angular momentum in rotating disks could be better
understood through the comparison of theoretical models and simulations
with observations of distant galaxies. In a first attempt at estimating
the angular momenta of distant galaxies, \citet{FS06} studied several
z$\sim$2 galaxies, and found that, $\sim$10 Gyr ago, galaxies appear to
have approximately the same specific angular momentum as today's spirals,
with values roughly similar to that expected for their haloes \citep[see
also][]{nesvadba06}. They argued that this confirms the hypothesis that
baryons likely acquired their angular momentum during the collapse of
their parent dark matter halos. However, as they pointed out, the spatial
resolution of their observations does not allow to uniquely distinguish
between rotating disks and merger-induced kinematics, thus the origin
of high angular momentum in z$\sim$2 disk galaxies still remains unclear.

At lower redshift, a sample of z$\sim$0.6 galaxies has been observed
using the multi-integral field spectrograph FLAMES/GIRAFFE at VLT
\cite[see] [hereafter Paper I and Paper II, respectively]{flores06,
puech06}. GIRAFFE observations are confronted with similar
difficulties in term of spatial resolution as any other study of high
redshift galaxy dynamics. However, to mitigate against these effects,
we developed a classification scheme based on the kinematics and
morphologies of the galaxies, separating them into rotating disks and
galaxies with complex or disturbed kinematics and morphologies (see
Paper I for details). Given all of the other comparisons made in
subsequent papers, e.g., Paper II and \citet[][hereafter Paper
III]{puech06b}, this classification method appears very robust (see
detailed discussions in both Paper I and Paper II). The goal of this
paper is to derive the specific angular momentum in these z$\sim$0.6
galaxies. The kinematics and emission line properties of this sample
has been studied in detail in the three previous papers of this series
(dedicated to the GIRAFFE Guaranteed Time Observation sample; see
Paper I, Paper II, and Paper III). This paper is the forth of this
series and is organized as follows: \S 2 summarizes the observations
and introduce the methodology. \S 3 presents the specific angular
momentum of the GIRAFFE sample. \S 4 discusses the dynamical support
of distant rotating disks, while \S 5 and 6 discuss the implications
and summarize our results.

\section{Observations \& Methodology}

As part of the Guaranteed Time Observations (GTO) of the Paris
Observatory, we obtained observations with the multi-integral field
spectrograph FLAMES/GIRAFFE, of a complete sample of 32 galaxies,
with redshifts ranging from 0.4 to 0.7, EW$_o$(OII)$\ge$ 15\AA$\;$ and
I$_{AB}\le$22.5. Briefly, we used the LR04 and LR05 grating, changing
the setups to specifically target the [OII]$\lambda\lambda$3726,3729
doublet (with R\(\sim10000\)); integration times ranged from 8 to 13
hours; the seeing was typically \(\sim 0.6-0.7\) arcsec during all the
observations. Data cubes were reduced using the GIRBLDRS v1.12 package
\citep{blecha00}, including a flat-fielding. Sky was carefully subtracted
with our own IDL procedures. We derived for these galaxies both velocity
fields and velocity dispersion maps in Paper I and Paper II. These maps,
as well as HST morphology, were used to divide the sample into three
distinct classes based mainly on their dynamical characteristics: rotating
disks, perturbed rotators, and galaxies with complex kinematics. A
complete description of the GTO sample and of the methods and analysis
used to classify the galaxies from our sample are given in Paper I.

The specific angular momentum $j$ of a rotating system can be
estimated as \citep[see, e.g.,][]{mo98}: $$j=\beta R V_{max},$$ where
$\beta$ is a dimensionless parameter that depends on the geometry and
spatial distribution of mass, $R$ a characteristic radius of the mass
distribution, and V$_{max}$ is the maximal rotation velocity. For a
thin exponential disk of scale length $R_d$, this relation becomes:
$$j_{disk}=2R_d V_{max}.$$

\begin{figure}[h!]
\centering
\includegraphics[width=9cm]{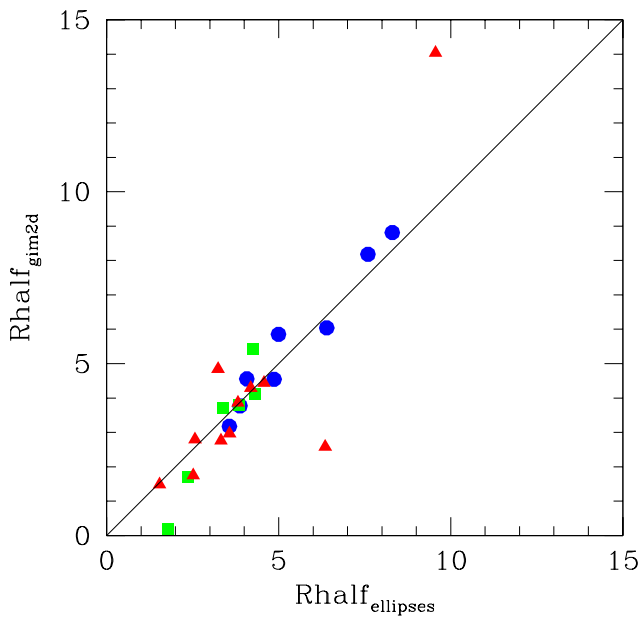}
\caption{Comparison of half-light radii (in kpc) derived using GIM2D
and using the modeled results from fitting isophotal ellipses to the
galaxy light profile. The resulting half light radii derived from both
methods agree very well, with the galaxies with disturbed or complex
kinematics showing additional scatter. {\it Blue dots} represent
galaxies classified as rotating disks, {\it green squares} represent
perturbed rotations, and {\it red triangles} represent galaxies with a
complex kinematics.}
\label{Fig1}
\end{figure}

To estimate a disk scale length, $R_d$, it is necessary to deconvolve
the disk from the bulge component. We used GIM2D \citep{simard98,
simard02} to measure $R_d$ in the galaxies observed with GIRAFFE using
HST/WFPC2 images (0.1 arcsec pixel$^{-1}$) or with preference, ACS
images (0.05 arcsec pixel$^{-1}$) when they were available. In Paper I
and Paper II, we derived the half light radius from the modeling of
isophotal ellipses \citep[see also][]{hammer01}. Figure \ref{Fig1}
shows that both half light radii derived using this method and GIM2D
agree very well. Unfortunately, GIM2D does not estimate any
uncertainty for the half light radius, but it is noteworthy that a
linear fit, between the half light radii derived using GIM2D and the
modelling of isophots, returns a median standard deviation of
$\sim$0.29 kpc, in close agreement with the typical uncertainty of
$\sim$0.34 kpc on the half light radius derived from the isophotal
ellipses modelling, as claimed in Paper I. Given the fitting methods
are independent, this comparison suggests that our R$_d$ measurements
using GIM2D are robust. Uncertainties on $R_d$ were directly taken as
1-sigma uncertainties returned by GIM2D, with a median value of 0.12
kpc.

\begin{figure}[h!]
\centering
\includegraphics[width=9cm]{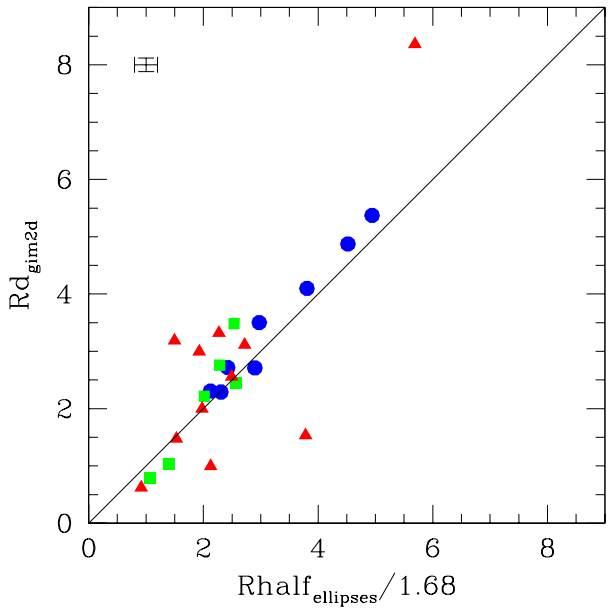}
\caption{Comparison of disk scale lengths (in kpc) derived using GIM2D
and using the modeled results from fitting isophotal ellipses to the
galaxy light profile. The resulting disk scale length from both
methods agree very well, with the galaxies with disturbed or complex
kinematics showing additional scatter. {\it Blue dots} represent
galaxies classified as rotating disks, {\it green squares} represent
perturbed rotations, and {\it red triangles} represent galaxies with a
complex kinematics. The median uncertainties, i.e., 0.12 kpc for
GIM2D, and 0.34/1.68 for the modelling of the isophots respectively,
are indicated in the upper-left corner.}
\label{Fig2}
\end{figure}

Unfortunately, for five galaxies, we only had ground-based images
available which were taken at the CFHT \citep[see, e.g.,][; 0.207
arcsec pixel$^{-1}$]{schade96}. Because these images are significantly
affected by the relatively large seeing disk (large compared to the
$R_d$ of the disks), we did not attempt to use GIM2D to fit their disk
light profiles since certainly the solution would be highly
degenerate, and thus unlikely to be reliable. Instead, we used the
fact that for an exponential disk, $R_d$ can theoretically be obtained
from the half light radius $r_{\rm half}$ using $R_{d}$=$r_{\rm
half}$/1.68. In Figure \ref{Fig2}, we compared $R_d$ as measured by
GIM2D with $r_{\rm half}$/1.68 as deduced from Paper I and Paper II.
This figure shows that this is a reasonable assumption (at least for
rotating quiescent disk galaxies). We also used this method for two
galaxies for which GIM2D failed to provide a statistically robust fit
to the light profiles.

Both Figures \ref{Fig1} and \ref{Fig2} show that for galaxies
exhibiting more complex kinematics than a simple rotating disk (i.e.,
those classified as perturbed rotation or complex), the estimates of
$R_d$ and/or $r_{\rm half}$ have a larger relative dispersion. In
general, however, even the galaxies with complex kinematics have $R_d$
and/or $r_{\rm half}$ consistent with the average results of the
rotating disks. The relatively high dispersion can be easily
understood as a result of their unrelaxed dynamical state: it is
likely that these galaxies are not in rotational equilibrium, and an
exponential disk model is then probably an inadequate representation
of their true light profile. However, to get homogeneous and
consistent estimates, we choose to treat the whole sample, whatever
the kinematical class is, as if all galaxies were rotating disks. We
also note that despite the increased scatter between the various
estimates of the scale length and half-light radius, each kinematic
subsample generally falls along the one-to-one line. It is simply that
the individual estimates are less reliable for the galaxies with
complex kinematics but not for the ensemble of each class.

$V_ {max}$ was estimated by fitting a double gaussian to the [OII]
doublet (see Paper I). This introduces a relative random uncertainty
on the measurement of $V_{max}$ that is estimated to be $\sim$ 10
km/s, from the comparison of independent fits to the data. This
uncertainty is relatively low because of the high spectral resolution
of GIRAFFE (R$\sim$10000) which allows a good velocity measurement.
Moreover, the limited size of the IFU can introduce an observational
source of uncertainty, if this size is too small to reach the flat
part of rotations curves. However we showed in Paper I that the size
of the GIRAFFE IFU is well-suited to measure $V_{max}$ for all the
galaxies of the sample.

Because of the coarse sampling of the GIRAFFE IFU data (pixel size of
0.52 arcsec, $\sim$ 3.5 kpc at z$\sim$0.6), a mean correction of 20\%
has to be applied on $V_{max}$. This correction factor is appropriate
for z$\sim$0.6 rotating disks (see discussion of this in Paper I).
This correction factor was determined by simulating GIRAFFE
observations from hydrodynamical simulations of a Sbc-like rotating
disk. As explained in Paper I and Paper II, we are probably severely
underestimating this correction factor for objects showing more
complex kinematics (see Paper II), and this could introduce a relative
offset between rotating disks and objects with more complex
kinematics. The absolute uncertainty of the correction factor for rotating
disks was determined to be $\pm$4\% because of possible variations in
size and inclination (see Paper I). An additional source of uncertainty on
this correction factor could be due to variations of the dynamical
properties along, e.g., the Hubble sequence. We checked that, using
another hydrodynamical simulation of a ``mean'' SDSS galaxy, a mean
20\% correction factor is still appropriate, within $\pm \sim$2\%
relatively to 20\% correction factor. We emphasize that this uncertainty is
very difficult to estimate, given the high number of parameters which
have to be taken into account.


Finally, uncertainties on $j_{disk}$ were estimated using usual
methods, i.e. propagating the individual uncertainties on $V_{max}$,
$R_d$, and on the correction factor on $V_{max}$, as detailed above.
This gives a median 1-sigma uncertainty of 0.09 dex and 0.10 dex in
$V_{max}$ and $j_{disk}$, respectively. This is to be compared with
typical uncertainties for local galaxies of 0.03 and 0.06 dex
respectively, as estimated from \citet{courteau97}. We emphasize that
such an uncertainty is meaningful only for rotating disks, since the
accuracy and appropriateness of the model used to reproduce the data
is not taken into account. This uncertainty should then be viewed as a
random uncertainty associated with $j_{disk}$, if the rotating disk
model is correct, but does certainly not include the systematic
uncertainty when such a model is not appropriate. Indeed, in Papers I
and II, we emphasized the fact that objects showing complex kinematics
are probably mergers or merger remnants. For these objects, the
$j_{disk}$ derived as above should then be viewed as the
\emph{orbital} angular momentum due to the relative motion of the two
progenitors, rather than the intrinsic \emph{spin} angular momentum of
a single rotating disk \citep[see also][]{FS06}. We will discuss this
in more detail subsequently.

\section{Angular Momentum of Distant Galaxies}


To compare distant with local galaxies, we used two local samples from
\citet{courteau97} and \citet{mathewson92}. For the first sample, we
used the compilation made by \citet{steinmetz99}, and kindly provided
by M. Steinmetz. For the second sample, we retrieved the
electronically available data from the
CDS\footnote{http://cdsweb.u-strasbg.fr/}. For this sample, no $R_d$
measurement were directly available: we derived it from their (I band)
23.5 magnitude isophotal radius dividing by a mean correction factor
of 3.5 \citep{palunas00}.

\begin{figure}[h!]
\centering
\includegraphics[width=9cm]{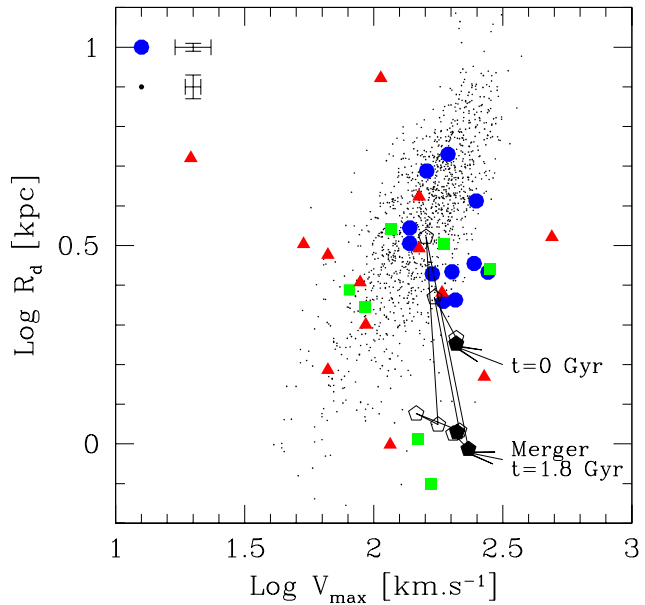}
\caption{Disk scale length versus maximal rotational velocity, for the
sample of 32 distant galaxies presented in Paper I and Paper II. {\it
Blue dots} represent galaxies classified as rotating disks, {\it green
squares} those classified as perturbed rotators, and {\it red
triangles} galaxies with complex kinematics. The {\it small black
dots} represent the local sample of \citet{courteau97} and
\citet{mathewson92}. The median 1-sigma uncertainties are indicated in
the upper-left corner, for both distant rotating disks (blue dots),
and local disks (black dots). For local galaxies, $R_d$ have been
determined using 1D fits whereas 2D fits were used for distant
galaxies: this explains why the uncertainty along the ordinate is
larger for local than for distant galaxies. Black pentagons represent
simulated GIRAFFE observations using a hydrodynamical simulation of a
major merger of two Sbc galaxies. The beginning and the end of the
sequence, as well as the merger itself, are indicated by solid
symbols.}
\label{FigRV}
\end{figure}

Figure \ref{FigRV} shows the disk scale length versus the maximal
rotation velocity, both for local and distant samples. Both local
samples agree quite well, in spite of the different proxies used for
the maximal rotational velocity and the disk scale length. Distant
rotating disks fall close to local rotating disks, although some of
them appear to have a relatively lower disk scale length. Those are
associated to galaxies with half light radius slightly lower than that
of compact galaxies (i.e., $R_{half} \leq$ 4.7 kpc, see Paper II). On
the other hand, more kinematically disturbed distant galaxies show a
very high dispersion around the local $R_d-V_{max}$ relation. It can
be explained in two ways. First, more kinematically complex distant
galaxies have a larger uncertainty in the determination of their
radius. This uncertainty can be estimated from the scatter of
kinematically complex galaxies in Figure \ref{Fig2}, and is $\sigma
\sim$1.45 kpc. Relatively to the median $R_d$ of kinematically complex
galaxies, this translates into a 0.2 dex scatter in Figure
\ref{FigRV}. It is thus clear that this is not sufficient to explain
the extremely large scatter of the kinematically complex galaxies in
the $R_d-V_{max}$ plane (i.e., $\sigma \sim$0.45 dex relatively to the
fit of the local galaxies $R_d-V_{max}$ relation). The second (and
most important) source of scatter is that rotational velocities
undergo abrupt spatial and temporal variations during dynamical
processes such as minor and/or major mergers. This effect is
illustrated by the black pentagons track: they represent a
hydrodynamical simulation of a merger of two Sbc galaxies from
\cite{cox06}. GIRAFFE observations were simulated along the
merging sequence, as well as HST images (Puech et al., in
preparation). From these simulated data, both rotation velocities and
disk scale lengths were extracted and corrected following similar
methods used for real distant galaxies (see previous section). During
this simulation, the merger occurs at $\sim$1.8 Gyr after the
beginning of the simulation, and is simulated for up to 0.5 Gyrs after
the merger. At the end of the simulation, the remnant looks relatively
similar to an elliptical, with a small inner disk at the center. It is
beyond the scope of this paper to explore for every possible track in
Figure \ref{FigRV}, varying, e.g., the gas fraction, orbital
parameters, or the amount of feedback. What is important to note here
is that this simulation, which includes possible observational or
instrumental biases, illustrates how the scatter of the most
kinematically disturbed distant galaxies could be easily reproduced by
such events. This strengthens the idea that these objects are probably
ongoing merger and/or merger remnants.

\begin{figure}[h!]
\centering
\includegraphics[width=9cm]{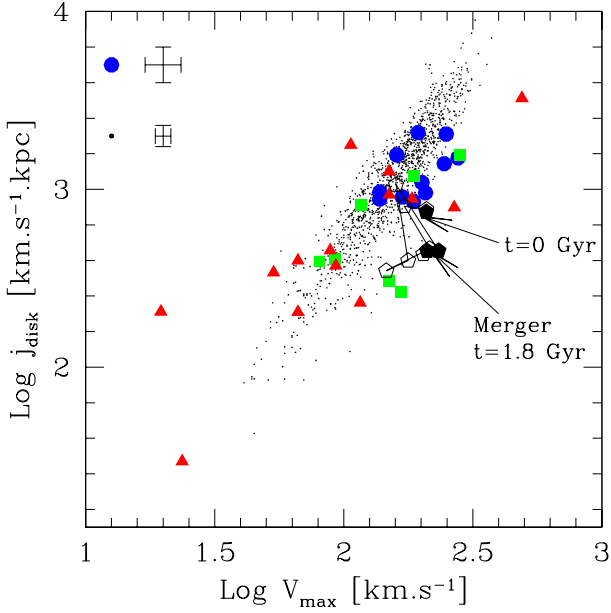}
\caption{Specific angular momentum of the disk, $j_{disk}$, versus
their maximum rotation velocity, V$_{max}$, for the sample of 32
distant galaxies presented in Paper I and Paper II. {\it Blue dots}
represent galaxies classified as rotating disks, {\it green squares}
those classified as perturbed rotators, and {\it red triangles}
galaxies with complex kinematics. The {\it small black dots} represent
the local sample of \citet{courteau97} and \citet{mathewson92}. The
median 1-sigma uncertainties are indicated in the upper-left corner,
for both distant rotating disks (blue dots), and local disks (black
dots). Black pentagons represent simulated GIRAFFE observations using
a hydrodynamical simulation of a major merger of two Sbc galaxies. The
beginning and the end of the sequence, as well as the merger itself,
are indicated by solid symbols.}
\label{Fig3}
\end{figure}

Figure \ref{Fig3} shows the specific angular momentum of the rotating
disks in the sample of galaxies observed with GIRAFFE, $j_{disk}$,
versus their maximum rotation velocity, V$_{max}$. In this figure,
z$\sim$0.6 rotating disks show a specific angular momentum comparable
to that of local galaxies with same $V_{max}$ (i.e., $\sim$ same total
mass). 
Objects with a complex kinematics have a lower median specific
angular momentum, with $log j_{disk} \sim 2.7$, than in rotating
disks, which have a median of $log j_{disk} \sim 3.0$. This is
probably again due to the fact that $V_{max}$ can be severely
underestimated for galaxies with complex kinematics (see Paper I and
Paper II, and above): their derived specific angular momentum should
then be considered as lower bounds. As in Figure \ref{FigRV}, the
dispersion of distant galaxies with complex kinematics is
significantly higher. As these objects are hypothesized to be
undergoing a merger or are merger remnants, we are likely measuring an
orbital angular momentum rather than an intrinsic spin angular
momentum of a single disk. In such a case, the shape factor $\beta$=2
is probably not appropriate, and could be an source of additional
dispersion. To investigate this effect, we also over-plotted in this
Figure the same simulation of GIRAFFE observations of a merger of Sbc
galaxies, as described above. Note that the beginning of the sequence
corresponds to only one of the two progenitors, which shows a deficit
in specific angular momentum compared to local galaxies. This could be
related to the difficulty for hydrodynamical simulations to reproduce
the higher angular momentum observed in local disks. The position of
most of galaxies with complex kinematics in the $j_{disk} - V_{max}$
plane can here again be easily explained as a result of mergers (see
black pentagons). As these simulated data include many possible
sources of uncertainty, it clearly suggests that the larger scatter of
galaxies showing complex kinematics can be associated with mergers or
merger remnants.

\section{Dynamical Support of Distant Disk Galaxies}

In this section, we compare the dynamical support (rotation vs.
velocity dispersion) of distant and local rotating disks. Such an
analysis could help us to understand how rotating disks acquire and
lose their angular momentum.

The most revealing quantity to estimate the dynamical support of
galaxies is the ratio of rotation velocity to velocity dispersion of
the disk, $V/\sigma$. $V$ is a circular velocity (quantifying the
amount of rotation), and $\sigma$ is an estimate of the intrinsic
velocity dispersion in the disk (i.e., turbulent or peculiar motions).
In elliptical galaxies, $V/\sigma$ is usually estimated via the ratio
of the maximal rotational velocity to the mean velocity dispersion
within 0.5$r_e$, where $r_e$ is the effective radius of the system
\citep[e.g.,][] {davies83, bender94}. For spiral galaxies, there is no
general consensus: different spatial components of $\sigma$, estimated
following different methods, have been used \citep[e.g.,][]{bottema93,
vegaBeltran01, hunter05}. \citet{binney05} demonstrated that with
integral field spectroscopy (i.e., a 2-dimensional spatial coverage
combined with simultaneous spectral coverage), the intrinsic
$V/\sigma$ is more robustly estimated using the ratio between the mean
squared rotational velocity and the mean squared velocity dispersion,
both measured directly along the line-of-sight.

We choose to estimate $V$ using the maximal rotational velocity
(corrected from inclination and spatial resolution effect, see \S~2),
as it is probably the most accurate quantity derivable from GIRAFFE
velocity fields. Because the velocity gradient of the rotation curve
of $z\sim$0.6 rotating disk galaxies falls approximatively in only one
GIRAFFE IFU pixel (or about one resolution element given the seeing,
see \S 2), the center of the GIRAFFE velocity dispersion maps show a
peak that is due to shear and/or large-scale motions in velocity, and
cannot be used to estimate the intrinsic velocity dispersion of the
disk (see Paper I and Paper II). On the contrary, the regions
surrounding the peak of the velocity dispersion are much less affected
by the shear and/or large-scale motions in velocity, since the
rotation curve is approximatively flat in these regions (i.e., has a
constant velocity with radius). These outer regions of the velocity
dispersion maps can then be used to construct a reliable estimate of
the intrinsic velocity dispersion of the disk. To construct such
reliable estimates of the intrinsic velocity dispersion of the disks,
we first removed the $\sigma$ peak due to rotation in the velocity
dispersion map, guided by the modeling of the dispersion we made in
Paper I. We then estimated $\sigma$ by deriving the signal-to-noise
weighted mean of the remaining pixels.

Unfortunately, there are very few published velocity dispersion maps
of local galaxies that could be used for direct comparison. However,
several dozen velocity dispersion profiles of the gas in local disks
that were obtained with long-slit spectroscopy have been published. We
combined the samples of \citet{vegaBeltran01}, \citet{corsini03}, and
\citet{pizzella04}, which are composed of spiral galaxies with
morphological type earlier than Sc. Note that $V/\sigma$ ratio does
not seem to depend on the morphological type, since a similar range of
values are found for both spiral and irregular galaxies
\citep[see][]{hunter05}. We checked that both local and distant
samples have similar distribution in the absolute B magnitude, $M_B$,
and we kept only galaxies with absolute B magnitude brighter than the
lower value found in the GIRAFFE sample (i.e., $M_B(AB) \leq$ -19.26),
to ensure that both samples are probing galaxies with comparable
stellar masses and star-formation rates. To exclude the central
dynamically hot region, we performed a sigma clipping on the velocity
dispersion profile of local galaxies, keeping only the points along
the curve that were below 2$\sigma$ around the median. We then took
the final median of the remaining points as a measure of $\sigma$ in
the disk.

When measured projected onto inclined disks, the observed velocity
dispersion is a combination of the three spatial components of the
velocity dispersion -- the radial component $\sigma _r$, the azimuthal
component $\sigma _{\phi}$, and the vertical component $\sigma _z$
\citep[e.g.,][]{binney89}: $$\sigma _{obs}^2=(\sigma_r^2
\sin^2{\eta}+\sigma _{\phi}^2 \cos^2{\eta})\sin^2{i}+\sigma _z^2
\cos^2{i}$$ where $\eta$ is the angle between the observed PA and the
major axis of the galaxy, and $i$ is the inclination angle of the
disk. In local spiral galaxies, $\sigma _z \sim \sigma _{\phi} \sim
0.7 \sigma _r$ for stars \citep[e.g., ][]{hunter05}. In the case of
local galaxies, if we assume both a well aligned slit and that gas and
stars dynamics are well coupled \citep[see][]{vegaBeltran01,
pizzella04}, one gets $\sigma _{obs} \sim \sigma _z$. In the case of
distant galaxies, observed using integral field spectroscopy, it is
necessary to correct for the different $\eta$ corresponding to the
different IFU pixels. We can directly correct for this effect on the
mean sigma by averaging the $\cos^2{\eta}$ and $\sin^2{\eta}$ terms,
and then multiplying by a $(1+\sin^2{i}/2)^{-1/2}$ correction factor.
Finally, we obtain an estimate of the spatially averaged $\sigma _z$,
that can be used in the $V/\sigma$ ratio. To check that the final
estimate of $V/\sigma$ does not depend significantly on the method
used to estimate $\sigma$, we compared with the velocity dispersion
measured at $r_{\rm half}$/4 as given by \citeauthor{vegaBeltran01}.
Using this alternative method does not significantly affect the
general trend of $V/\sigma$ estimated in local disks.

In the sample of local galaxies, errors have been estimated as
follows. The uncertainty on $V$ was directly taken as the mean of the
velocity measurement uncertainty on $V_{max}$ and $V_{min}$ as given
by \citet{vegaBeltran01}, \citet{corsini03}, and \citet{pizzella04}.
The uncertainty on $\sigma$ was estimated as the median of measurement
uncertainties of the points of the velocity dispersion curve kept
during the sigma clipping (see above). A conservative uncertainty of
three degrees was assumed for the inclination. We found a median
uncertainty of 1.6. Figure~\ref{Fig7} shows how this uncertainty
evolves with $V/\sigma$. This Figure shows that high values of
$V/\sigma$ have large associated uncertainties; including only sources
with $V/\sigma$ estimates greater than three times their uncertainty,
left us with a range of local $V/\sigma$ values ranging between 2 and
9, with a median uncertainty in $V/\sigma$ of 1.1. In the following,
we will consider these points only, as others galaxies have $V/\sigma$
ratios that are too uncertain. For distant rotating disks, we
estimated the uncertainty of $\sigma$ to be lower than 15 km/s (see
Paper I). We added an additional uncertainty term due to the
correction factor used to deconvolve $V_{max}$ (see \S~2).

\begin{figure}[h!]
\centering
\includegraphics[width=9cm]{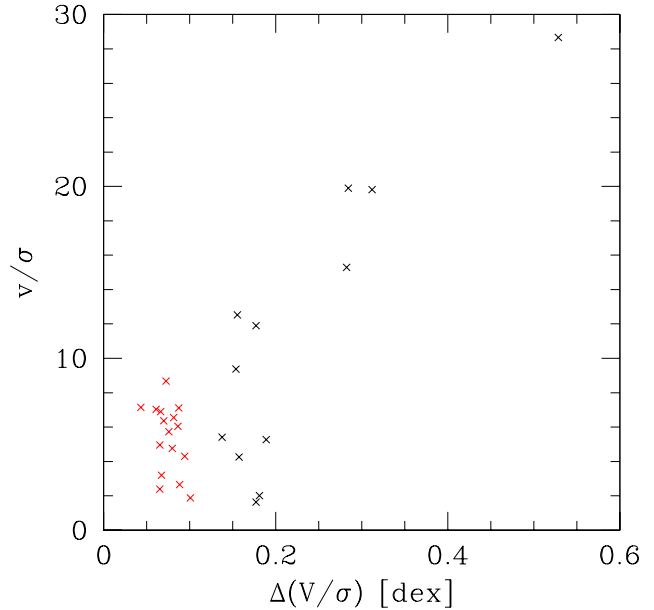}
\caption{$V/\sigma$ vs. uncertainty in $V/\sigma$ for the sample of
local galaxies used for comparison. The galaxies with significant
$V/\sigma$ estimates ($V/\sigma \geq$ 3$\Delta$ ($V/\sigma$)) are
shown in red.}
\label{Fig7}
\end{figure}

\begin{figure}[h!]
\centering
\includegraphics[width=9cm]{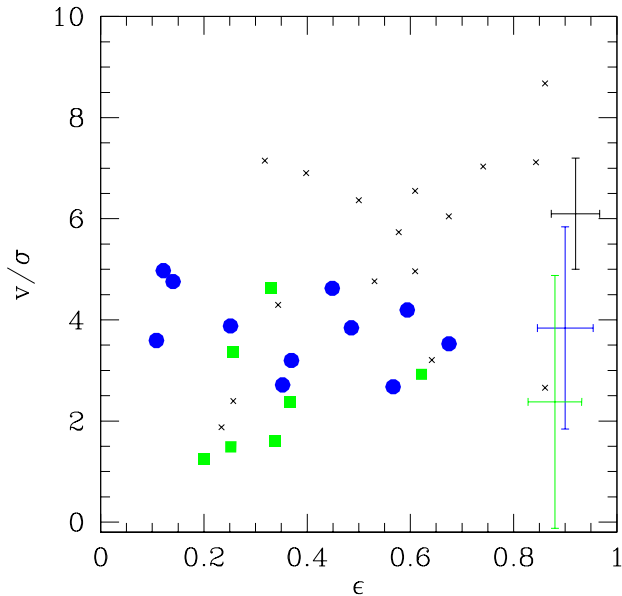}
\caption{Comparison of $V/\sigma$ and ellipticity, $\epsilon$, for
galaxies at z$\sim$0.6 compared to a matched sample of local galaxies.
The $V_{max}/\sigma$ ratio for galaxies at z$\sim 0.6$ that have been
determined to be rotating disks (blue dots) and to have rotation
curves that are perturbed (green squares) are compared with local
galaxies whose morphology and velocity curves are consistent with
rotating disks (black crosses). The medians are indicated by the
position of the median uncertainties on the right side of the diagram
(each being represented by matching colors)}.
\label{Fig6}
\end{figure}

Figure~\ref{Fig6} shows the comparison of $V/\sigma$ as a function of
the ellipticity in both distant and local rotating disk galaxies. The
median $V/\sigma$ is 3.8$\pm$2 for distant rotating disks, and
6.1$\pm$1.1 for local disks. Recall that highest local $V/\sigma$ have
been discarded in this comparison because of a too high uncertainty:
this could result in an underestimation of the local median
$V/\sigma$. We also plotted the $V/\sigma$ ratio for the distant
perturbed rotating disks, with a median of 2.4$\pm$2.5. Figure
\ref{Fig6} shows that distant disks have lower $V/\sigma$ ratios than
local ones. This is consistent with the fact these galaxies are likely
undergoing a minor merger and/or a gas accretion event which is
heating their disks \citep{walker96, velazquez99, abadi03}, as was
suggested in Paper II. More data will be needed to confirm this trend,
however.

\section{Discussion}

\subsection{The Growth of Angular Momentum in Galaxies with Time}

We showed in \S~3 that z$\sim$0.6 and local disk of same total mass
show comparable specific angular momentum. More dynamically complex
objects fall around the same $j_{disk}-V_{max}$ relation but with much
larger dispersion than the local relation. How can we interpret this
in terms of angular momentum growth with time in galaxies ?

One of the first theories proposed for the build up of angular
momentum in galaxies is the tidal torque theory \citep{peebles69,
white84}. The basic idea in the tidal torque theory is that most of
the angular momentum is being gradually acquired in the linear regime
of growth of the density fluctuations, due to tidal torques from
neighboring fluctuations. This process continues relatively
efficiently until the halo reaches its turn around time, i.e., roughly
when the protogalaxy decouples from the Hubble flow. Moreover, this
theory suggests that angular momentum gain is minimal when haloes are
growing non-linearly after they have decoupled from the expanding
background and have formed virialized systems. This theory then
predicts rapid growth of angular momentum in the early evolution of
massive halos, and subsequently little growth \citep{porciani02}. At
first sight, our results, combined with those of \citet{FS06} and
\citet{nesvadba06} would mean that the specific angular momentum of
rotating disks does not evolve with redshift from z$\sim$2 to present.
This would indicate that this scenario for the growth of angular
momentum is plausible. However, we have to caution that current
samples are relatively small and incomplete in terms of galaxy mass.
Moreover, there are some issues with the physical resolution of the
data, especially at z$>$2 \citep[but see][]{nesvadba06}. Even though
these results are all formally consistent with no overall evolution in
the specific angular momentum, given the limitations of all studies,
it is difficult to conclude anything about evolutionary trends in the
specific angular momentum in rotating disk galaxies.


Beyond this, the results presented here can constrain how angular
momentum and indeed how galaxies grow after the epoch when massive
halos (like the ones studies here) acquired their angular momenta
through tidal torquing. The basic ideas that have emerged from
theoretical research are that the specific angular momentum can be
increased or decreased in a halo through a random walk process of
interactions and merging with other halos \citep[e.g.,][]{gardner01,
vitvitska02, porciani02} or through gas accretion onto the halo
\citep[e.g.,][]{white78}. 
In the merger process, angular momentum is gained or lost depending on
the geometry, dynamics, the detailed mass distribution of the merging
halos, strength of the feedback, the relative masses of the merging
halos and galaxies, etc. \citep[e.g.,][]{robertson05, springel05,
d'Onghia06}. Alternatively, the angular momentum of disk galaxies
could grow through the accretion of cooling halo gas or smooth
accretion of mass into the dark matter halo \citep[e.g.,][]{white78,
peirani04, vandenBosch01, vandenBosch02a, vandenBosch02b,
vandenBosch03, chen03, okamoto05}. The amount of angular momentum
change then depends on the relative angular momentum of the gas and
dark matter, how the gas was accreted, how much mechanical dissipation
the gas undergoes during collapse, the orientation of its angular
momentum vector relative to the perhaps pre-existing disk, etc.

Returning to both Figs.~\ref{FigRV} and \ref{Fig3}, it is clear that
perturbed rotation and kinematically complex galaxies are
\emph{dynamically} different from dynamically relaxed local disks. The
dispersions in the $j_{disk}$-$V_{max}$ are indeed similar for the
rotating disks, higher for the perturbed rotators, and very high for
the galaxies with complex dynamics.
We can also note that some of the distant galaxies which are not
rotating disks have offsets to both lower and higher $j_{disk}$
relative to local galaxies for the same $V_{max}$. Because we are
likely underestimating $j_{disk}$ in kinematically complex galaxies
(see \S~3.1), we cannot conclude specifically about any difference on
median $j_{disk}$ between the different kinematical classes. However,
the higher dispersion in the $V_{max}$-$j_{disk}$ plane of
kinematically complex galaxies relatively to distant rotating disks is
significant. This dispersion is caused by abrupt variations of the
angular momentum in galaxies with complex kinematics, and are likely
the result of a random walk during a non-linear phase of evolution of
the angular momentum. This is the first observational evidence for
such a non-linear growth of the angular momentum in galaxies, as
expected from theoretical models \citep[e.g.,][]{gardner01,
vitvitska02, peirani04}.


An important question is what is the driving mechanism of this random
walk. Theoretical models show that major mergers cause the most abrupt
variations of the angular momentum, while minor mergers and/or gas
accretion are associated with smoother variations
\citep[e.g.,][]{vitvitska02, peirani04, hetznecker06}. If the gas is
heated to approximately the virial temperature of the halo before
collapsing to form or onto a pre-existing disk \citep{dekel06}, we
would expect the gas to have a specific angular momentum similar or
greater than that of the dark matter halo
\citep[e.g.,][]{chen03,okamoto05}. This is why we might then expect
that the specific angular momentum of the disk to remain constant or
to mildly increase or decrease at late times \citep{peirani04}. It is
hard to believe that smooth gas accretion alone could explain the
complex or perturbed kinematics and/or discrepant values of the
specific angular momentum.
We instead suggest that mergers (both major and minor mergers,
possibly associated with complex kinematic galaxies, and perturbed
rotations, respectively, see Paper I and II) play a significant role
in changing the angular momentum of galaxies with time. While overall
consistent with the tidal torque theory, the increased dispersion in
specific angular momentum and spin of the perturbed rotators and
galaxies with complex kinematics is consistent with the merger
scenario. \citet{vitvitska02} show that within the context of merging,
we would find both increases and decreases in the specific angular
momentum, resulting in an increase of the dispersion of their angular
momenta. This is consistent with what we observe when we compare
distant rotating disks to galaxies with complex kinematics. This
interpretation is strongly supported by a simulation of GIRAFFE
observations of a major merger. This simulation illustrates how the
higher dispersion of kinematically complex galaxies can arise from
mergers, taking into account all possible observational uncertainties
.

\subsection{The Building of Local Disk Galaxies}
In the following discussion, we will assume that most of the observed
distant galaxies with emission lines are progenitors of local disks.
This should be the case even for most galaxies showing complex
kinematics and perturbed rotations. Indeed, those galaxies represent
40\% of galaxies at z$\sim$0.6, and if they were E/S0 progenitors,
there would be a much higher fraction of E/S0 that it is observed
today (see, e.g., Hammer et al. 2005, and also Lotz et al. 2006).

Because we are comparing samples of galaxies spanning different total
mass ($V_{max}$), linking distant to local disks from Figure
\ref{Fig3} alone is not straight-forward. Can z$\sim$0.6 galaxies
evolve towards local disks through major mergers? \citet{hammer05}
claim that 75\% of local spiral in the $10^{10.5}$-$10^{11.5}M_\odot$
range \citep[the so-called ``intermediate-mass'' range, see
also][]{hammer06} have experienced a major merger since z=1. Assuming
an evolution rate as (1+z)$^{2.7}$ \citep{lefevre00}, one can derive
that $\sim$ 29\% of local spiral galaxies have experienced a major
merger since z=0.6. Similarly, \citet{lotz06} estimated that between
33\% and 66\% of $L_B \ge 0.4 L_B^*$ galaxies had a major merger since
z=1.1 with an evolution rate of (1+z)$^{1.12}$. Combining the above
estimates implies that $\sim$ 15 to 30\% of local spirals could have
experienced a major merger since z=0.6. Then, the majority of
z$\sim$0.6 rotating disks cannot evolve towards local spirals through
major mergers. Another possibility, is that z$\sim$0.6 galaxies could
evolved to z=0 spirals through minor mergers and/or gas accretion:
their rotating disks would then have survived, and their specific
angular momentum should statistically increase by only $\sim$ 0.1 dex
\citep{peirani04}. A third possibility is that distant disks would
evolve in a ``closed box''. In absence of external torques, their
specific angular momentum would then remain constant. Finally, the
last two possible evolution tracks from z$\sim$0.6 to z=0 are equally
viable.

This picture is also supported by the fact that distant disks (we
consider here both rotating disks and perturbed rotating disks
together) seem to be heated relatively to local disks (i.e., have
systematically higher $V/\sigma$), possibly through minor mergers
and/or gas accretion \citep{walker96, velazquez99, abadi03}. A
difficulty to quantify uncertainty in this result could arise from the
difference observational strategies used in the local and distant
samples (slit vs. integral field spectroscopy, respectively). This,
however, is unlikely given the fact that most spirals today are
strongly dominated by rotational motions \citep[e.g.,][]{binney89}. An
interesting additional component of the evolution in $V_{max}/\sigma$
we observed is the possibility of ``angular momentum mixing'' during
minor mergers and/or gas accretion events \citep[see][]{okamoto05}.
During such events, the direction of the angular momentum vector can
change, and newly accreted gas may then settle to a different
orientation from the pre-existing disk. During this process, the
observational manifestations are likely to be relatively high velocity
dispersion of the gas caused by the gas being shock heated by
turbulence generated by the overlapping and dynamically different
disks (what we could observe in Figure \ref{Fig6}).

Given the above discussion (see \S~5.1) about the specific angular
momentum, it is unlikely that (smooth) gas accretion alone plays an
important role in the general growth of rotating disks. It is also
equally unlikely that gas accretion alone could explain the disturbed
kinematics of perturbed rotation, as this process is expected to be
dynamically smooth (see Paper I and II) and only lead to small changes
in the angular momentum. Note, however, that, to our knowledge, there
is no clear theoretical prediction about the kinematics of gas
accretion by galaxies. Thus, our results show that it is likely that
mergers (and, \emph{a priori}, both major and minor mergers) play an
important role in the general build-up of rotating disks. The link
between dynamical processes such as mergers, and the dynamical
classifications adopted in this series of paper will be addressed in a
forthcoming article.

The results presented in this paper fit quite nicely within the
``spiral rebuilding'' scenario proposed by \cite{hammer05}, where
$\sim$75$\pm25$\% of local spirals (those of early type) have
undergone a merger since z=1 and have rebuilt a disk thanks to gas
accretion. This scenario is composed of 3 major phases: a ``pre-merger
phase'' during which two distant spirals merge, the ``LCG phase''
where all material from the progenitors fall into the mass barycenter
of the system and form a bulge, and the ``disk growing phase'' where
subsequently accreted material forms a rotating disk \citep[see Fig. 8
of][for how this might look]{okamoto05}. As the two progenitors are
merging, their disks, and thus, their spin angular momentum, are
destroyed during the collision \citep{cox04}. At the same time, the
encounter has a significant quantity of orbital angular momentum which
can be progressively converted into spin angular momentum as a new
disk is re-built around the remnant. Moreover, we saw that the
increase of $V/\sigma$ in disk galaxies could be due to heating
mechanisms such as minor mergers and gas accretion events, as if some
of the disk-rebuilding were triggered by accretion of the gas left
over from the merging event.
\citet{yoachim06} suggest that the properties of thick and thin disks
are consistent with gas rich mergers playing a significant role in their
formation where the stars in these mergers formed the thick disk while
the settling gas formed much of the thin disk. The results presented here
amplify these ideas and show their credibility in directly explaining
the dynamics of intermediate redshift disk galaxies.

\section{Conclusions}

We have studied the angular momentum and the dynamical support of a
sample of z$\sim$0.6 galaxies observed with the integral field
spectrograph FLAMES/GIRAFFE. We found that the classification of
distant galaxies based on their kinematic properties (mainly) and
morphologies into three distinct classes, i.e., rotating disks,
perturbed rotators, and kinematically complex, is apparently robust.
This classification appears to also select galaxies with angular
momenta consistent with local spiral galaxies but show varying degrees
of dispersion relative to the local values, increasing from
rotating disks to kinematically complex galaxies.

This can be interpreted as an evidence for a non-linear random walk
evolution of the angular momentum during the last 8 Gyr. A natural
driver for this random walk is provided by major mergers, since the
dispersion of kinematically complex galaxies in the $j_{disk}-V_{max}$
plane, as well as the complexity of their velocity fields itself (see
Puech et al. 2007, in preparation), can both be reproduced by
hydrodynamical simulations of such events. Major mergers also explains
how the angular momentum of local rotating disks could be acquired and
be as high as observed, namely through the conversion of orbital
angular momentum to spin momentum. Moreover, distant disks appear to
be more turbulent (lower $V/\sigma$). This is likely an indication
that local disks could grow through the accretion of gas through
mergers or in discrete clouds. In a refinement of this general
picture, the higher velocity dispersions could be due to ``angular
momentum mixing'' \citep[see][]{okamoto05} whereby the relative orbit
of the infalling gas is skewed compared to the previous disk. The
torque provided by the accreting gas causes a change in the angular
momentum vector thereby increasing the dispersion observed in the gas.

Finally, these findings are consistent with the observational scenario
proposed by \citet{hammer05}, where $\sim$75\% of local spirals (those
of early type) have undergone a major merger since z=1, and have then
rebuilt their disks thanks to gas accretion (possibly from high
angular momentum gas left over from the merging event) and/or minor
mergers. Unfortunately, drawing robust conclusions as to the physical
processes driving the observational manifestations we have discussed
in this paper is difficult given the small numbers of galaxies we have
studied. To increase the statistical robustness of these results, we
are currently analyzing similar data for a much larger sample of
several hundred disk galaxies as part of the VLT Large Program IMAGES
(P.I.: F. Hammer). The galaxies are selected by absolute J-band
magnitude, to have redshifts of or less than 0.9, and to have [OII]
equivalent width comparable to the galaxies study here
\citep{ravikumar06}. With the robust selection and large numbers, we
should be able to make more definitive statements about the physical
behind the dynamical and morphological evolution of spiral galaxies
over the last 7 Gyrs.

\begin{acknowledgements}
We would like to thank M. Steinmetz for having providing us with their
compilation of data on the kinematics of local galaxies, and for
interesting comments on recent simulation work. We are especially
indebted to T.J. Cox who provided us with hydrodynamical simulations
of a Sbc galaxy. We also would like to thank C. Balkowski, P. Amram,
and L. Chemin for very interesting discussions we had about the
subject of this paper.
\end{acknowledgements}

\appendix

\section{Spin Parameter of the Haloes}
We present in this appendix a perhaps more predictive comparison
between local and distant galaxies' angular momentum, based on the use
of the spin parameter of the surrounding haloes, which has the
advantage of being approximately independent of the halo mass
\citep{barnes87, lemson99, maller02a, peirani04}. This dimension-less
parameter measures the ratio between the true angular momentum of the
system and the angular momentum it would have, if all the mass was
entirely supported by rotation, or, equivalently, the ratio of the
spin energy to the total binding energy. The spin parameter
\citep{peebles69} is given by: $$ \lambda =
\frac{J|E|^{1/2}}{GM^{5/2}}, $$ where $J$ is the true halo angular
momentum, $E$ the total energy and $M$ the total mass of the system.
It is possible to estimate $\lambda$ using observed characteristics of
the disk. Following the method outlined by \citet{hernandez06},
$\lambda$ can be estimated using: $$\lambda \sim R_d / V_d^2.$$ This
formulation requires the following assumptions: (1) $j_{disk} \sim
j_{halo}$ (see discussion in the introduction); (2) the total disk
mass must be a constant fraction of the mass of the halo;
(3) the total energy is dominated by the halo which is virialized; (4)
a baryonic Tully-Fisher relation exists which can be expressed by $M_d
\sim V_d^4$ \citep[see][]{mcgaugh05}\footnote{\citet{hernandez06} used
  a baryonic Tully-Fisher relation slightly different of the one used
  in this paper, leading them to $\lambda \sim R_d/V_d^{3.5}$.}. We
additionally require for our analysis to be valid that these
assumptions are not dependent on the epoch of observation. In Figure
\ref{Fig7}, we plotted the histogram of $\lambda$ in the local sample
used in this paper. The dash line at $\lambda \sim$ 0.03 is the most
probable local value found by \citet{tonini06} using an independent
method. It is very close the peak of our histogram: in spite of its
crudeness, this suggests that the method to estimate $\lambda$ is
appropriate, at least for local disks.

\begin{figure}[h!]
\centering
\includegraphics[width=9cm]{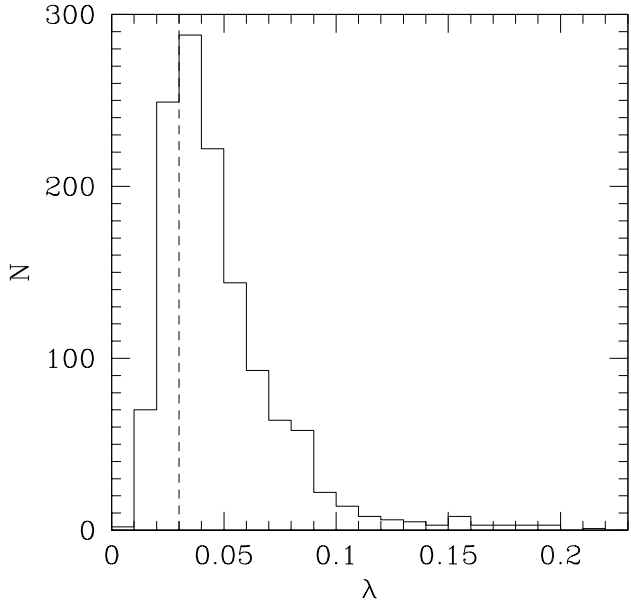}
\caption{Histogram of $\lambda$ derived using the method outlined in the
text for the local sample of galaxies. The most probable value is very
close to estimate made by \citet{tonini06} also using a local sample
of galaxies.}
\label{Fig7}
\end{figure}

\begin{figure}[h!]
\centering
\includegraphics[width=9cm]{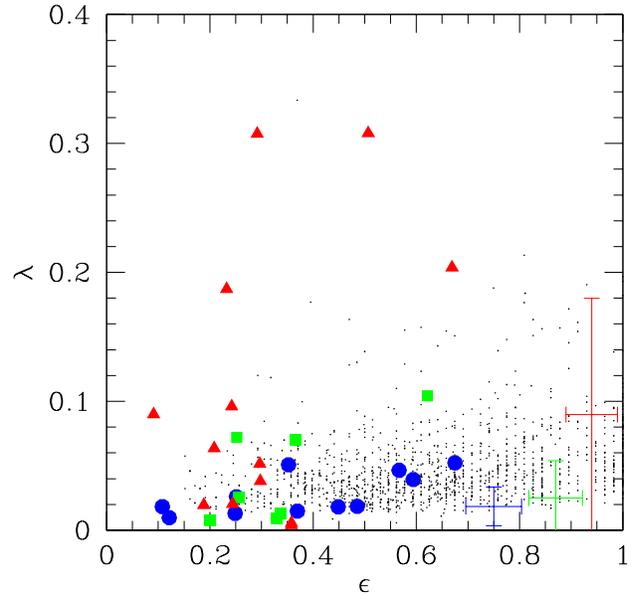}
\caption{Spin parameter, $\lambda$, vs. ellipticity, $\epsilon$, in
local and distant galaxies samples. {\it Blue dots} represent galaxies
classified as rotating disks, {\it green squares} represent those
classified as perturbed rotators, and {\it red triangles} represent
galaxies with complex kinematics. {\it Black dots} represent the local
sample of \citet{courteau97} and \citet{mathewson92}. The medians are
indicated by the position of the median uncertainties for distant
galaxies are indicated on the right side of the plot (each being
represented by matching colors). Errors include both standard
propagation of errors, and a bootstrap estimation of the error
associated with the determination of the median itself.}
\label{Fig5}
\end{figure}

In Figure \ref{Fig5}, we used this relation to investigate the
difference in the spin parameter, $\lambda$, between local and distant
galaxies. From \citet{peirani04}, the median $\lambda$ between z=0.6
and z=0 is expected to remain roughly constant with accretion only,
whereas an increase of $\sim$10\% is expected, taking into account
major merger events. Figure \ref{Fig5} seems to favor a small increase
in $\lambda$ from z$\sim$0.6 to z=0. Unfortunately, the uncertainty on
the median $\lambda$ of the z$\sim$0.6 galaxies is very high, see Fig.
\ref{Fig5}: it is then not possible to identify the relative
contribution of these two mechanisms. Both perturbed rotations and
objects with complex kinematics show higher dispersion in the spin
parameters than rotating disks (both local and distant). Part of this
dispersion is probably due to the rotation velocity being
under-estimated for the galaxies which are not dynamically relaxed
(see above). It could be tempting to claim for an increasing
dispersion among the three dynamical classes, from rotating disks to
objects with complex kinematics, as was seen in the specific angular
momentum, in agreement with a random walk model driven by mergers.
However, uncertainty also increases from rotating disks to
kinematically complex galaxies, and it becomes then very difficult to
firmly claim for such an evolution in the dispersions. Although it
seems possible to measure $\lambda$ for local galaxies, its derivation
seems still very difficult for more distant objects.

\end{document}